\documentclass[journal]{IEEEtran}

\usepackage{cite}
\usepackage{amsmath,amssymb,amsfonts}
\usepackage{algorithmic}
\usepackage{graphicx}
\usepackage{textcomp}
\usepackage{xcolor,url}
\usepackage{nomencl}

\begin{document}

\title{Incomplete Air Mixing Reduces the Efficiency of Commercial Buildings Behaving as Virtual Batteries}

\author{\IEEEauthorblockN{Austin J. Lin, Jacques A. de Chalendar,  and Johanna L. Mathieu, \textit{Senior Member, IEEE}}
\thanks{This work is supported by NSF GRFP DGE 2241144. A.J. Lin and J.L. Mathieu are with the Department of Electrical
Engineering and Computer Science, University of Michigan, Ann Arbor, MI
48109 USA (e-mail: \{aulin, jlmath\}@umich.edu). J.A. de Chalendar is with the Department of Energy Science and Engineering, Stanford University, Palo Alto, CA 94305 USA (e-mail: jdechalendar@stanford.edu).}}

\maketitle

\makenomenclature

\begin{abstract}
    Commercial building Heating, Ventilation, and Air Conditioning (HVAC) systems can provide flexibility to the electricity grid. Some researchers have found it convenient to model HVAC systems as virtual batteries. These models also better align with models used by grid planners and operators. However, experiments have shown that HVAC load shifting can be inefficient, and virtual battery models do not capture this inefficiency well. While the models typically use the average room temperature as the system's ``state of charge," they do not capture other factors that affect HVAC power/energy such as airflow and mixing. Here, we develop a new analytical building model to explore how incomplete mixing of supply air into a conditioned space leads to inefficiency in a virtual battery capturing the dynamics of HVAC fan power load shifting. The model qualitatively matches experimental results better than previous models, and shows that, as mixing becomes worse, the virtual battery becomes less efficient. Unfortunately, air mixing is unmeasured/unmeasurable. However, we show that, by closing the loop around measurements of fan power, we can improve the virtual battery's performance without the need for air mixing measurements. For example, in one case, we show a roundtrip efficiency improvement from 0.75 to 0.99.     
\end{abstract}


\nomenclature{\(t_\mathrm{start}\)}{Time event starts.}
\nomenclature{\(t_\mathrm{end}\)}{Time event ends.}
\nomenclature{\(t_\mathrm{settle}\)}{Time building is assumed to have settled.}
\nomenclature{\(E_\mathrm{in}\)}{Energy used to charge virtual battery.}
\nomenclature{\(E_\mathrm{out}\)}{Energy discharged by virtual battery.}
\nomenclature{\(P_\mathrm{f}\)}{Fan power.}
\nomenclature{\(\text{RTE}\)}{Round trip efficiency of virtual battery.}
\nomenclature{\(\alpha\)}{Tolerance for energy-neutral events.}
\nomenclature{\(T_\mathrm{r}\)}{Room temperature.}
\nomenclature{\(T_\mathrm{r,base}\)}{Baseline room temperature.}
\nomenclature{\(Q_\mathrm{sa}\)}{Supply air heat gain.}
\nomenclature{\(Q\)}{Internal heat gain.}
\nomenclature{\(C_\mathrm{r}\)}{Thermal capacitance of room.}
\nomenclature{\(R\)}{Thermal resistance of walls.}
\nomenclature{\(C_\mathrm{w}\)}{Thermal capacitance of walls.}
\nomenclature{\(T_\mathrm{w}\)}{Wall temperature.}
\nomenclature{\(T_\mathrm{oa}\)}{Outside air temperature.}
\nomenclature{\(\dot{m}_\mathrm{sa}\)}{Supply airflow rate.}
\nomenclature{\(C_\mathrm{p,a}\)}{Specific heat of air.}
\nomenclature{\(T_\mathrm{sa}\)}{Supply air temperature.}
\nomenclature{\(T_\mathrm{a}\)}{Mixing zone temperature.}
\nomenclature{\(R_\mathrm{a}\)}{Thermal resistance of mixing zone.}
\nomenclature{\(C_\mathrm{a}\)}{Thermal capacitance of mixing zone.}
\nomenclature{\(c\)}{Relative thermal capacitance of mixing zone.}
\nomenclature{\(r\)}{Relative thermal resistance of mixing zone.}
\nomenclature{\(T_\mathrm{set}\)}{Temperature setpoint.}
\nomenclature{\(\tau_\mathrm{sa}\)}{Supply airflow time constant.}
\nomenclature{\(\tau_\mathrm{fan}\)}{Fan power time constant.}
\nomenclature{\(\beta\)}{Fan power coefficient.}
\nomenclature{\(k_\mathrm{p,temp}\)}{Temperature control proportional gain.}
\nomenclature{\(k_\mathrm{i,temp}\)}{Temperature control integral gain.}
\nomenclature{\(k_\mathrm{p,pow}\)}{Power control proportional gain.}
\nomenclature{\(k_\mathrm{i,pow}\)}{Power control integral gain.}
\nomenclature{\(\dot{m}_\mathrm{d,sa}\)}{Desired supply airflow rate.}
\nomenclature{\(P_\mathrm{f,diff}\)}{Difference between measured and baseline fan power.}

\printnomenclature

\section{Introduction}

As the penetration of intermittently-available renewable generation increases, there is a growing opportunity for loads to actively participate in balancing supply and demand~\cite{TAYLOR_GRIDS_WITHOUT_FUEL}. Commercial building heating, ventilation, and air conditioning (HVAC) systems have been identified as potential candidates for active loads~\cite{ZHAO2013225, Callaway_controllable_loads, AustinPESGM, Wang_Virtual_Batt}. The large thermal mass of buildings allows for HVAC power to be modulated without causing immediate changes to building temperature or occupant comfort, making the HVAC load flexible~\cite{Adi_ERIS}. With more intermittent renewable generation, the need to balance loads at sub-hourly timescales will become crucial to grid stability~\cite{Makarov_windgen_intrahour}. Commercial building HVAC systems, specifically HVAC fans, have demonstrated the ability to perform demand response at these timescales~\cite{AustinPESGM, Adi_ERIS, BEIL_EFF_FAST_DR, Hao_HVAC}. 

It is convenient to model active loads as ``virtual batteries," to make deployment among the wide variety of flexible resources similar~\cite{KATS_VB, Hughes_Virtual_batteries, Abbas_virtual_batteries_waterheaters, Hao_VB}. The battery analogy has been applied to HVAC fans providing demand response~\cite{Wang_Virtual_Batt, Hao_VB, Hughes_Virtual_batteries}, where load consumed above the counterfactual baseline (i.e., what the fans would have otherwise consumed) is akin to charging the battery while load consumed below the baseline is akin to discharging the battery. When an HVAC system is in cooling mode, ``charging'' the building makes it colder while ``discharging'' makes it warmer~\cite{BEIL_EFF_FAST_DR, Raman_RTE1, Adi_ERIS}. The ``battery state of charge" is usually assumed to be the average room temperature~\cite{Raman_RTE1}.

Past results have found that virtual batteries capturing the dynamics of HVAC fan power during sub-hourly load shifting are inefficient, which, in some cases, means that buildings consume more energy than they would have otherwise~\cite{AustinPESGM, Lin_2R2C_model_paper, Raman_RTE1, Adi2019}. In other cases, buildings consume less energy than they would have otherwise, which might indicate an impact to other building services that affect occupant comfort, e.g., temperature and ventilation~\cite{Adi_ERIS, AustinACEEE}. There is also an efficiency gap between experimental results and building model simulation results, where simulations tend to predict better efficiency metrics~\cite{Lin_2R2C_model_paper, AustinPESGM, Adi_ERIS, Raman_RTE1}. Previous simulation results indicated that inefficiency was a byproduct of restoring the average room temperature back to the setpoint value at the conclusion of a load shifting event~\cite{Lin_2R2C_model_paper, Raman_RTE1}. These simulation results disagree with experimental results~\cite{Adi_ERIS, AustinACEEE}, specifically on the characteristics of the post-event settling of the fan power trajectory. Further, the experimental results do not show a clear link between the fan power and average room temperature, as expected by the virtual battery analogy~\cite{austin_asme}.

It remains unclear how inefficient building load shifting actually is, why the inefficiencies exist, and what building systems compensate for the change in energy consumption. It is also unclear the extent to which the inefficiency is inherent to the building versus how much can be reduced through better or closed-loop control. The building automation and control system may also be missing critical measurements that explain the inefficiency. To achieve a better understanding of the efficiency of virtual batteries, we need better models, sensing, and associated analysis.

In this paper, we show that variables not typically monitored by building automation and control systems can negatively impact the performance of virtual batteries capturing the dynamics of HVAC fan power load shifting. We develop, analyze, and simulate a new analytical building model to explore how incomplete mixing of supply air into the conditioned space impacts the virtual battery and its efficiency. The new model qualitatively matches experimental results better than previous simulation models. We find that, as mixing becomes worse, the virtual battery becomes less efficient. We also find that past results have assumed buildings return to normal operation after a load shifting event more quickly than is likely reasonable, causing the perceived inefficiency to be worse than the actual inefficiency. 

Given these results, it would seem valuable to monitor and incorporate variables for air mixing into load shifting strategies to improve building load shifting efficiency. However, we find that closed-loop control of fan power can significantly improve load shifting efficiency without the need for air mixing measurements, which, in any case, would be difficult/impossible to obtain in practice. Specifically, we develop a controller that  dynamically adjusts the temperature setpoint during the events instead of using predetermined temperature setpoints fully-specified offline. This strategy treats the building and its existing controllers as a black box and closes the loop around real-time fan power consumption measurements, which are easier to obtain than air mixing measurements. This solution is also easier to scale than solutions that modify existing building control loops~\cite{LivingLabs}. 

The contributions of this paper are 1) a new analytical building model and associated analysis, 2) demonstration of how unmeasured building phenomenon (e.g., air mixing) may negatively impact the performance of virtual batteries capturing the dynamics of HVAC fan power load shifting, and 3) simulation results showing how directly closing the loop around fan power measurements may counteract this. While our model focuses on air mixing, there may be other unknown and/or unmeasured sources of load shifting inefficiency, some of which might be inherent to the design of the building or its controllers. It is unrealistic and infeasible to measure all such variables; however, our closed loop control results show that improved virtual battery performance may be achievable without these measurements. This finding highlights the need for direct and real-time measurements of HVAC fan power consumption.

The rest of this paper is organized as follows. In Section~\ref{sec:background}, we describe how HVAC systems performing load shifting can act as virtual batteries. In Section~\ref{sec:model}, we introduce a new analytical building model including air mixing, the controllers used for load shifting, and the metrics used to analyze load shifting events. In Section~\ref{sec:results}, we provide our simulation results and, in Section~\ref{sec:conclusion}, we conclude the paper.

\section{HVAC load shifting as a Virtual Battery}
\label{sec:background}

In this section, we discuss the value and limitations of representing HVAC fan power load shifting as a virtual battery. We provide a brief background on typical HVAC operation, describe how load shifting events are generated, discuss the value and limitations of the  battery analogy in capturing the dynamics of HVAC fan power load shifting, and describe how insufficient air mixing may be a source of inefficiency.

\subsection{Typical HVAC operation}
\begin{figure}
    \centering
    \includegraphics[width=\linewidth]{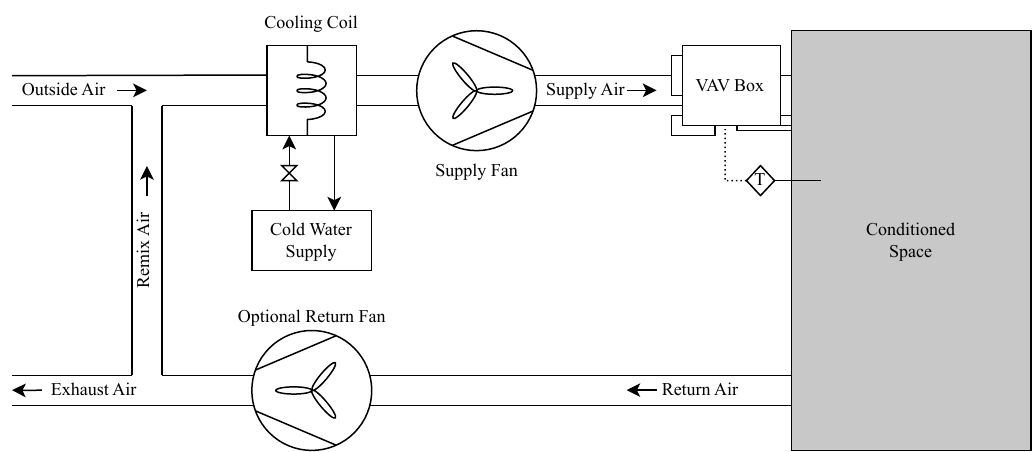}
    \caption{Illustration of the basic operation of a VAV HVAC system operating in cooling mode.}
    \label{fig:HVACDiagram}
\end{figure}

The objective of an HVAC system is to maintain temperature and ventilation throughout a building, which is done by moving air through ducts. This work focuses on variable airflow volume (VAV) HVAC systems, which represented 30\% of US commercial floor space in 2018~\cite{CBECS}. An illustration of a VAV HVAC system in cooling mode is shown in Fig.~\ref{fig:HVACDiagram}. Outside air is drawn into the ducts, mixed with remix air, cooled by a cooling coil, and forced through the ducts by the supply fan. The conditioned air is now referred to as supply air. The flow of cold water into the cooling coil is regulated to maintain a constant supply air temperature. The airflow volume of supply air that enters the conditioned space from the ducts is controlled by a VAV box, which uses a Proportional-Integral (PI) controller to regulate the room temperature to approximately match the temperature setpoint. If the room is warmer than desired, the VAV box increases airflow and vice versa. Air (referred to as return air) exits the conditioned space through the return duct, which may (but not always) include a return fan. Fan speed is controlled with a PI controller to maintain an adequate duct pressure to force air into and out of the room. To reduce energy consumption, some of the return air can be recycled back into the system as remix air (through use of an economizer). The rest of the return air is exhausted outside. The illustration in Fig.~\ref{fig:HVACDiagram} is simplified to show a single VAV box. In reality, a supply fan feeds multiple VAV boxes servicing multiple zones. For more information about HVAC operation, we refer the reader to~\cite{janis2009mechanical}.

\subsection{Generating load shifting events}
\label{sec:events}

The purpose of load shifting is to change when energy is consumed, not to reduce the overall consumption. On sub-hourly timescales, the HVAC chiller, which supplies cold water to the cooling coils, may not respond fast enough~\cite{Adi_ERIS}, so power flexibility is primarily provided by the HVAC fans. Fan power can be directly driven through control of airflow~\cite{Wang_Virtual_Batt}, which requires careful consideration of building constraints, such as maintaining room temperature and ventilation requirements. Fan power can be indirectly changed through changes in the temperature setpoint of all/most of the rooms, a process known as Global Thermostat Adjustment (GTA)~\cite{BEIL_EFF_FAST_DR, Adi2019, AustinPESGM, SHIFDR}. The advantages of GTA are that 1) it is much easier to implement than direct fan power control because it only requires changes to building control parameters (temperature setpoints) versus the control design itself, 2) GTA does not affect critical building functions, such as providing minimum ventilation requirements, since it preserves existing controls and overrides, ensuring occupant safety, and 3) occupant comfort is guaranteed if temperature setpoint changes are small. 

\begin{figure}
    \centering
    \includegraphics[width=\linewidth]{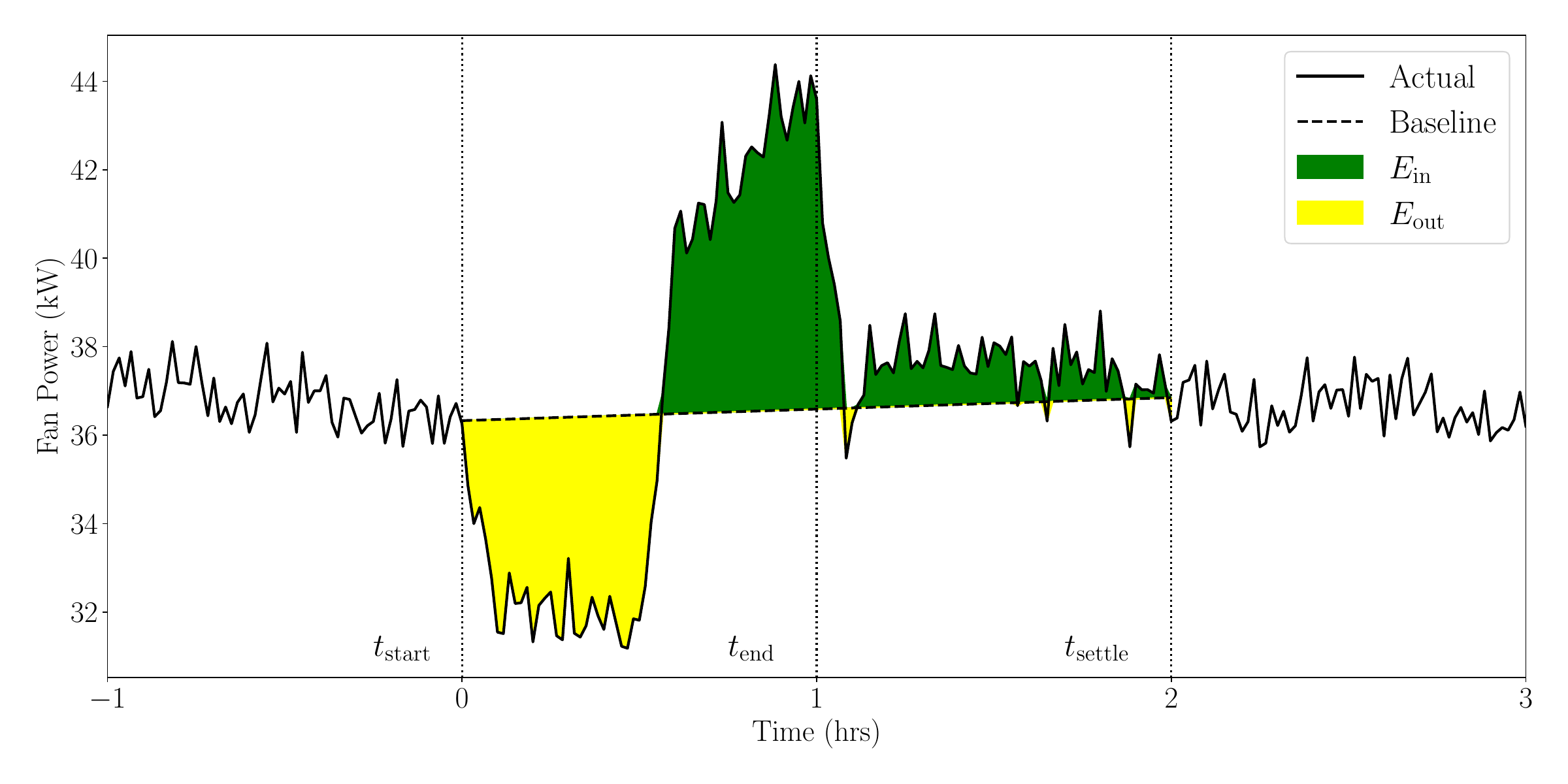}
    \caption{Example of the fan power during a DOWN-UP load shifting event. The actual fan power is shown compared to the baseline fan power. Time $t_\mathrm{start}$ is the start of the event, $t_\mathrm{end}$ is the end of the event, and $t_\mathrm{settle}$ is the point at which the building is assumed to have returned to normal operation. We also shade the regions corresponding to the input energy $E_\mathrm{in}$ and output energy $E_\mathrm{out}$, which are used in the battery model.}
    \label{fig:ExamplePower}
\end{figure}

An example of a GTA load shifting event in a real building is shown in Fig.~\ref{fig:ExamplePower} (using data from the Sub-metered HVAC Implemented For Demand Response (SHIFDR) dataset~\cite{SHIFDR}). To generate this event, the temperature setpoint is first increased to cause a decrease in fan power, then decreased to cause a increase in fan power, and then returned to normal operation. We call this a DOWN-UP event. An UP-DOWN event is when power is first increased and then decreased. In this paper, we only consider 1 hour long UP-DOWN and DOWN-UP events where temperature setpoint changes every 30 minutes. The baseline fan power (dashed line in Fig.~\ref{fig:ExamplePower}) is the estimated power consumption of the fans if no event had occurred. It is used to quantify the change in energy/power due to the event. There are a variety of common ways to estimate baselines; in Fig.~\ref{fig:ExamplePower} we use the linear baseline method described in~\cite{Shunbo_linear_baseline}. The difference between the baseline fan power and the actual fan power is the power trajectory of the virtual battery. The energy consumed above the baseline  $E_\mathrm{in}$ (green area in Fig.~\ref{fig:ExamplePower}) charges the virtual battery and energy consumed below the baseline $E_\mathrm{out}$ (yellow area in Fig.~\ref{fig:ExamplePower}) discharges it. These particular load shifting events were designed to enable estimation of the efficiency of buildings as virtual batteries~\cite{BEIL_EFF_FAST_DR}. They are not meant to emulate any specific type of demand response event or demand response program. However, the load-shifting timescale is consistent with that of (non)spinning reserve or real-time energy markets. 

\subsection{Value and limitations of the battery analogy} 

Modeling a building as a virtual battery simplifies its (complex) dynamics and provides estimates of its power and energy flexibility, enabling power system operators to easily incorporate these models/constraints within their planning and operational problems~\cite{Hughes_Virtual_batteries, Scharnhorst_VB_commBuildings_risk, Hao_VB}. 
As a result, there has been significant research into multiple aspects of virtual battery models including system identification, incorporation of uncertainty, optimization, and assessment of performance. For example, in~\cite{Hao_VB}, the authors identify a virtual battery model to fit both commercial buildings and battery energy storage systems, and use that within an optimal coordination approach. The authors of~\cite{Hughes_Virtual_batteries} generalize the virtual battery model from residential HVAC to fit commercial HVAC, while the authors of~\cite{Arafat_VB_waterheaters} extend a virtual battery model to include water heaters. In~\cite{Scharnhorst_VB_commBuildings_risk}, the virtual battery parameters are determined from building data, where risk is considered in the building flexibility. The authors of~\cite{Saberi_VB_estimation_risk} develop a method to manage uncertainty when deploying buildings modeled with virtual batteries for frequency regulation. In~\cite{Abbas_virtual_batteries_waterheaters} the performance of commercial building virtual batteries are compared to water heater virtual batteries in resource scheduling problems. The authors find that battery models perform similarly to complex thermal models for energy maximizing, energy minimizing, and power reference tracking scenarios. 

Another line of work on virtual batteries is to explore their inefficiency. Preliminary experimental studies found an average efficiency of 0.46 when considering only UP-DOWN events~\cite{BEIL_EFF_FAST_DR}. Subsequent model-based simulations designed to understand the causes of the inefficiency found efficiency values ranging from 0.85 to 1.17 (efficiencies larger than unity indicate the building consumed less energy than in the baseline case), and deviations in energy were required to restore the room temperature after the end of the event~\cite{Lin_2R2C_model_paper}. Further work using a similar model found that sequential load shifting events resulted in efficiencies approaching unity~\cite{Raman_RTE1}. These simulation results do not agree with experimental results where inefficiency can range from 0.39 to 2.46~\cite{AustinPESGM, Adi2019}, with little impact on room temperatures~\cite{Adi_ERIS, AustinACEEE}. Additionally, the direction of load shifting (i.e., UP-DOWN versus DOWN-UP) plays a role in whether the virtual battery efficiency is larger or smaller than unity. Simulation studies have found that DOWN-UP events had average efficiencies smaller than unity while UP-DOWN events had average efficiencies larger than unity~\cite{Raman_RTE1,Lin_2R2C_model_paper}. The opposite was found in experimental studies~\cite{AustinPESGM, Adi_ERIS}, confusing the link between room temperature and virtual battery efficiency. Other proposed explanations for the virtual battery inefficiency have been poor baseline estimation methods~\cite{Shunbo_linear_baseline}, building controller limitations~\cite{austin_asme}, or building pressure fluctuations~\cite{AustinACEEE}. 

\subsection{Air mixing as a possible source of inefficiency}

Incomplete mixing of air in the conditioned space may be a source of virtual battery inefficiency. It is often assumed that air in the conditioned space is well mixed, e.g.,~\cite{Lin_2R2C_model_paper, Raman_RTE1}, which implies 1) there is instantaneous heat transfer between the supply air and the room air, 2) the thermostat accurately measures the (average) temperature of the room, and 3) the supply air achieves its maximum cooling potential. If the air is not well mixed, the temperature measured by the thermostat may not be representative of the actual (average) room conditions, and so the building controller may take unintended actions
creating longer than expected post-event settling behavior. Furthermore, incompletely mixed air will be removed from the room and so the supply air will not achieve its maximum cooling potential reducing the effectiveness of the fans in cooling the conditioned space. Effectively, this means that energy may be used inefficiently simply to push some of conditioned air into and out of the space, and through the ducts, with little heat transfer into the space. Longer than expected settling and poor fan cooling effectiveness are the key characteristics of virtual battery inefficiency~\cite{AustinACEEE, AustinPESGM}.

By relaxing the assumption of well mixed air, we would remove the assumption that the (average) room temperature can be measured and we could, in theory, model air mixing, room temperature gradients, and supply air cooling potential. However, it may not be feasible or practical to obtain the measurements needed for accurate models of the dynamics of air mixing. Additionally, there may be undetected errors within the building measurement system~\cite{VINDEL_grid_interactive_capabilities}. Many building automation and control systems use technology that is too slow to accurately track sub-hourly dynamics. Without the exact knowledge of and trust in building measurements, it may be impossible to accurately measure, model, and control all relevant aspects of building load shifting such as air mixing. 

Despite these challenges, we can develop a model that captures how incomplete air mixing affects virtual battery performance. The purpose of such a model is not to capture the dynamics of any specific building, but rather the underlying characteristics of virtual batteries in order to gain an understanding of how (typically unmodeled) building physics could affect virtual battery efficiency. We next propose this model.

\section{A model for incomplete air mixing }
\label{sec:model}

In this section, we develop a new analytical mixing air model that considers incomplete mixing of air in the conditioned space. We first describe the original model used by~\cite{Lin_2R2C_model_paper} to explore building load shifting inefficiency. We then detail the modifications that we made to that model to include air mixing. After that, we describe two control approaches used for generating load shifting events and, finally, we detail our method of analyzing those events. 

We note that we do not consider incomplete mixing to be the sole contributor to virtual battery inefficiency. Other factors, such as supply temperature reset or VAV reheat, may also contribute. The purpose of our simulations is to demonstrate how unmeasured building phenomenon can impact the performance of the virtual battery. We do not believe it is feasible or practicable to identify all factors that contribute to virtual battery inefficiency. Instead, we will turn to closed-loop control to mitigate inefficiency without explicitly modeling all factors contributing to it.

\subsection{Original model}

\begin{figure}
    \centering
    \includegraphics[width=\linewidth]{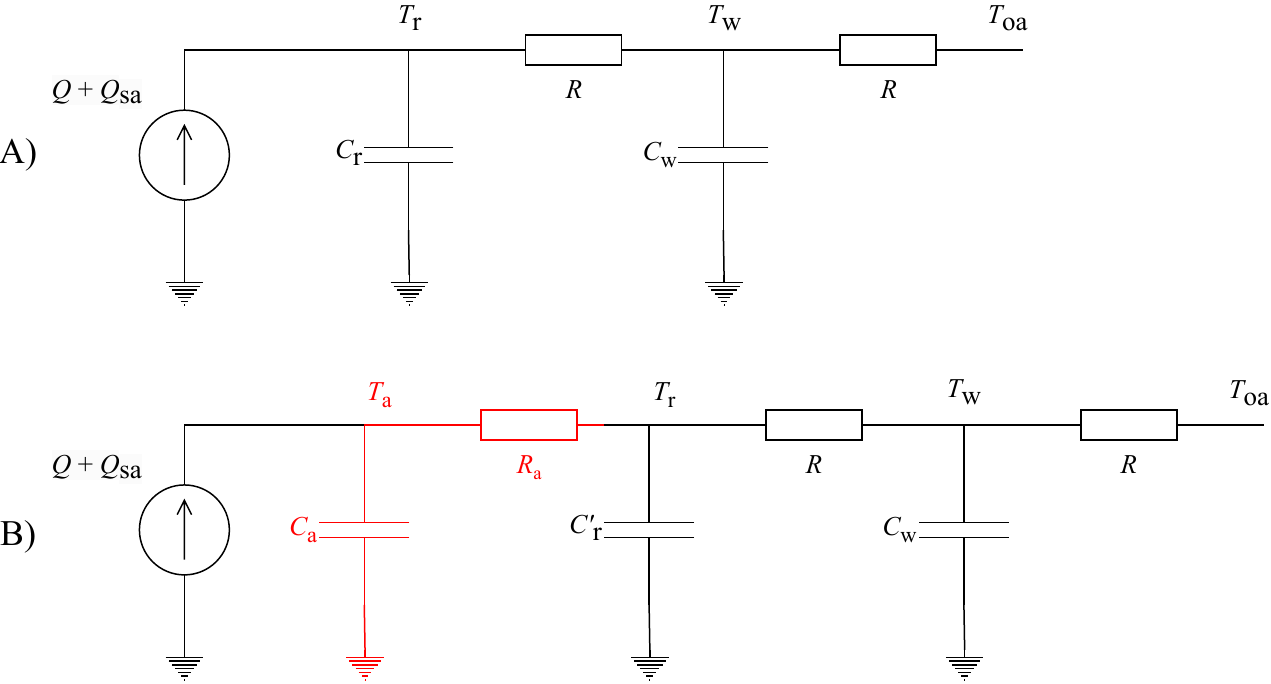}
    \caption{RC circuit model for (A) the model developed in~\cite{Lin_2R2C_model_paper} and (B) the new mixing air model developed in this paper. We highlight in red the addition of an extra state that represents air that has not been fully mixed into the room air and the associated thermal resistance and capacitance.}
    \label{fig:RCmodel}
\end{figure}

Our model is developed from the model used by Lin et al.~\cite{Lin_2R2C_model_paper}, which uses a standard resistance-capacitance (RC) network model for building thermal dynamics, shown in Fig.~\ref{fig:RCmodel}(A). RC thermal models equate the thermal resistance and capacitance of a building into an equivalent electrical circuit. Heat flow is akin to electrical current and temperature is akin to voltage. This modeling is done for heat conduction, not heat convection. The model used in~\cite{Lin_2R2C_model_paper}, has two resistors and two capacitors to represent the thermal resistances of the air-to-wall barrier and the thermal capacities of the rooms and walls. The walls in this model represent all thermal mass in the building that is not air. The heat transfer between individual rooms in the building is not considered.

Heat is added to the room through internal heat gain $Q$ and heat added by the supply air $Q_\mathrm{sa}$. The thermal capacity of the room is $C_\mathrm{r}$ and the room temperature is $T_\mathrm{r}$. The room air transfers heat to the walls with resistance $R$, representing the thermal insulation of the walls. The walls have a thermal capacitance $C_\mathrm{w}$ and temperature $T_\mathrm{w}$. The walls have a thermal resistance $R$ to the outside air, with temperature $T_\mathrm{oa}$. The air-to-wall resistance is assumed the same on both sides of the wall (inside and outside).

Room temperature is regulated by controlling the supply airflow, specifically, the mass flow rate of the supply air $\dot{m}_\mathrm{sa}$. The heat added by the supply air is calculated using the specific heat of the supply air
\begin{equation}
    Q_\mathrm{sa} = \dot{m}_\mathrm{sa}C_\mathrm{p,a}\left(T_\mathrm{sa} - T_\mathrm{r}\right),
    \label{eqn:spacificHeat}
\end{equation}
where $C_\mathrm{p,a}$ is the specific heat of air and $T_\mathrm{sa}$ is the supply air temperature. It is assumed that $T_\mathrm{sa}$ is constant. By applying \eqref{eqn:spacificHeat} to the model in Fig.~\ref{fig:RCmodel}(A), the dynamic equations are
\begin{eqnarray}
    C_\mathrm{r}\dot{T}_\mathrm{r} &=& \frac{1}{R}\left(T_\mathrm{w} - T_\mathrm{r}\right) + Q + \dot{m}_\mathrm{sa}C_\mathrm{p,a}\left(T_\mathrm{sa} - T_\mathrm{r}\right),\\
    C_\mathrm{w}\dot{T}_\mathrm{w} &=& \frac{1}{R}\left(T_\mathrm{r} - T_\mathrm{w}\right) + \frac{1}{R}\left(T_\mathrm{oa} - T_\mathrm{w}\right).
\end{eqnarray}
The parameter values for this model are listed in Table~\ref{tab:params}. These parameters are taken from~\cite{Lin_2R2C_model_paper} and calibrated to an auditorium building on the University of Florida campus.

\begin{table}
    \centering    
    \caption{Model parameters}
    \begin{tabular}{r|c|l}
    \hline
         Parameter & Variable &  Value\\
         \hline \hline
         Room thermal capacitance & $C_\mathrm{r}$ & $3.4 \times 10^7$ J/K\\
Wall thermal capacitance & $C_\mathrm{w}$ & $5.1 \times 10^7$ J/K\\
Wall thermal resistance &$R$ & 0.0013 K/W\\
Internal heat gain & $Q$ &	25 kW\\
Outdoor air temperature & $T_\mathrm{oa}$	& 85°F / 29.4°C\\
Supply air temperature & $T_\mathrm{sa}$ & 60°F / 15.6°C\\
Specific heat of air & $C_\mathrm{p,a}$ & 1000 J/kg-K\\
\hline
    \end{tabular}
    \label{tab:params}
\end{table}

\subsection{Mixing air model}
To model air mixing into the space, we add an intermediate state between the supply air and the room temperature, which we refer to as the mixing zone. Physically, the mixing zone represents a pocket of cold air around the supply duct outlet. The adapted RC network model is shown in Fig.~\ref{fig:RCmodel}(B), where $T_\mathrm{a}$ is the air temperature of the mixing zone, $R_\mathrm{a}$ is the thermal resistance between the mixing zone and the room air, and $C_\mathrm{a}$ is the thermal capacitance of the mixing zone. We also redefine the room capacitance as $C_\mathrm{r}'$, which accounts for some of the room air now existing in the mixing zone. All other model notation retain the same meaning as the original model. 

The supply air is now added to the mixing zone instead of the room air itself, i.e., 
\begin{equation}
    Q_\mathrm{sa} = \dot{m}_\mathrm{sa}C_\mathrm{p,a}\left(T_\mathrm{sa} - T_\mathrm{a}\right).
    \label{eqn:spacificHeatMixing}
\end{equation}
Larger values of $R_\mathrm{a}$ result in less heat transfer from the supply air to the room, and worse mixing. As the value of $C_\mathrm{a}$ increases, the amount of air in the mixing zone increases. In this scenario, the room thermostat may not provide a good measure of the actual conditions in the room.

Using Fig.~\ref{fig:RCmodel}(B) and \eqref{eqn:spacificHeatMixing}, the dynamics of the mixing air model are
\begin{eqnarray}
    C_\mathrm{a}\dot{T}_\mathrm{a} &=&\frac{1}{R_\mathrm{a}}\left(T_\mathrm{r} - T_\mathrm{a}\right) + Q + \dot{m}_\mathrm{sa}C_\mathrm{p,a}\left(T_\mathrm{sa} - T_\mathrm{a}\right),\label{eqn:model1}\\
    C'_\mathrm{r}\dot{T}_\mathrm{r}&=& \frac{1}{R_\mathrm{a}}\left(T_\mathrm{a} - T_\mathrm{r}\right) + \frac{1}{R}\left(T_\mathrm{w} - T_\mathrm{r}\right),\label{eqn:model2}\\
    C_\mathrm{w}\dot{T}_\mathrm{w}&=&\frac{1}{R}\left(T_\mathrm{r} - T_\mathrm{w}\right) + \frac{1}{R}\left(T_\mathrm{oa} - T_\mathrm{w}\right).
    \label{eqn:model3}
\end{eqnarray}
The parameter values are the same as the original model, with the following exceptions. Some of the room's thermal capacitance is now in the mixing zone. We define $c$ as the proportion of the conditioned space that the mixing zone occupies, and redefine the room capacitance and mixing zone capacitance as
\begin{eqnarray}
    C_\mathrm{a} &=& cC_\mathrm{r} ,\\
    C_\mathrm{r}' &=& (1-c)C_\mathrm{r},
\end{eqnarray}
which preserves the total thermal capacitance of the original model. Similarly, we define $r$ as the relative thermal resistance of the mixing zone compared to that of the walls
\begin{equation}
    R_\mathrm{a} = rR.
\end{equation}
The values of $r$ and $c$ represent relative sizes of the mixing zone resistance and capacitance and are changed during testing, as described in Section~\ref{sec:results}. In the limit $r=c=0$, the mixing zone disappears, and we recover the original model.

\subsection{Control}
\begin{figure}
    \centering
    \includegraphics[width=\linewidth]{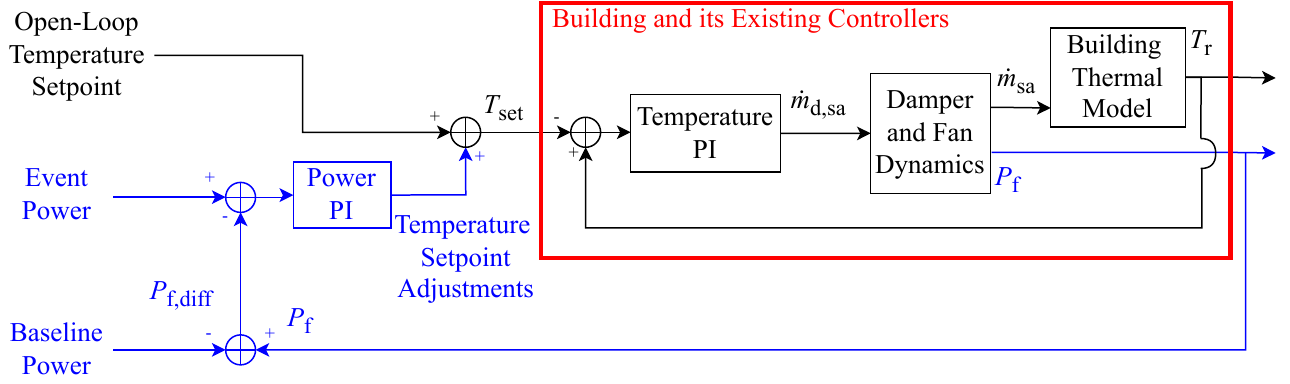}
    \caption{Block diagram of two control strategies: 1) open-loop load shifting via temperature setpoint control shown in black and 2) closed-loop load shifting via power control, which adds the loop shown in blue.}
    \label{fig:controldiagram}
\end{figure}

We consider two control approaches for generating load shifting events: temperature setpoint control and power control. Load shifting via temperature setpoint control leverages the controllers already implemented in most existing buildings. Under temperature setpoint control, room temperatures are regulated to maintain desired temperatures (setpoints). Open-loop load shifting can be achieved by increasing and decreasing the temperature setpoints according to predetermined schedules. In contrast, closed-loop load shifting can be achieved via power control, in which a new controller closes the loop around fan power measurements. Power control treats the building together with its existing temperature setpoint controllers as a black box and uses the mismatch between the measured fan power and the desired fan power to compute temperature setpoint changes to drive the measured fan power to the desired fan power. A block diagram is shown in Fig.~\ref{fig:controldiagram} and described in this section. The dynamics described in \eqref{eqn:model1}-\eqref{eqn:model3} are included in the block labeled ``Building Thermal Model." Table~\ref{tab:paramsControler} contains a summary of the control parameters. 

\begin{table}
    \centering    
    \caption{Controller parameters}
    \begin{tabular}{r|c|l}
    \hline
         Parameter & Variable &  Value\\
         \hline \hline
         Nominal Temperature Setpoint & $T_\mathrm{set}$ &	71°F / 21.7°C\\
         Supply Airflow Time Constant & $\tau_\mathrm{sa}$ & $30$ s\\
Fan Power Time Constant & $\tau_\mathrm{fan}$ & $150$ s\\
Fan Power Coefficient &$\beta$ & 220.8 W/(kg/s)$^3$\\
Temperature Control Proportional Gain & $k_\mathrm{p,temp}$ &	2\\
Temperature Control Integral Gain & $k_\mathrm{i,temp}$ &	0.001\\
Power Control Proportional Gain & $k_\mathrm{p,pow}$ &	3.33$\times 10^{-3}$\\
Power Control Integral Gain & $k_\mathrm{i,pow}$ &	2.083$\times 10^{-5}$\\
\hline
    \end{tabular}
    \label{tab:paramsControler}
\end{table}

Fan power, $P_\mathrm{f}$, is calculated using a linear relationship between fan power and airflow\footnote{There is disagreement on the best way to model this relationship; see~\cite{austin_asme}.}. Past results~\cite{austin_asme} have found that there is a small delay in the fan power response to changing airflow. We account for this by adding a low-pass filter between the airflow and fan power
\begin{equation}
    P_\mathrm{f}(s) = \frac{\beta}{\tau_\mathrm{fan} (s) + 1}\dot{m}_\mathrm{ma}(s),
\end{equation}
where $\beta$ is a linear coefficient and $\tau_\mathrm{fan}$ is the fan power time constant. This filter is included in the block labeled ``Damper and Fan Dynamics" in Fig.~\ref{fig:controldiagram}. We have picked values of $\tau_\mathrm{fan} = 150$~s and $\beta = 220.8$~W/(kg/s)$^3$ heuristically.

\subsubsection{Temperature setpoint control}
Temperature setpoint control captures how VAV boxes typically maintain room temperature. This controller is shown in the box labeled ``Temperature PI" in Fig.~\ref{fig:controldiagram}. The VAV box has a PI controller that tracks the error between room temperature and temperature setpoint to determine the desired airflow. Based on~\cite{Lin_2R2C_model_paper}, we use $k_\mathrm{p,temp} =2$ for the proportional gain and $k_\mathrm{i,temp}=0.001$ for the integral gain. The nominal temperature setpoint is 71°F (21.7°C).

In real buildings, there are physical limits to how quickly VAV boxes respond to control action. Like in~\cite{Lin_2R2C_model_paper}, we add a first-order low-pass filter between the desired supply airflow and the actual supply airflow, i.e.,
\begin{equation}
    \dot{m}_\mathrm{sa}(s) = \frac{1}{\tau_\mathrm{sa} s + 1} \dot{m}_\mathrm{d,sa}(s),
\end{equation}
where $\dot{m}_\mathrm{d,sa}$ is the desired supply airflow and $\tau_\mathrm{sa}$ is the supply airflow time constant. These dynamics are included in the block labeled ``Damper and Fan Dynamics" in Fig.~\ref{fig:controldiagram}. By adjusting the ``Open-Loop Temperature Setpoint," we can create open-loop load shifting events similar to past work~\cite{Adi2019, SHIFDR}, without the need for the power controller.

The building control system presented in this section is simplified to model the basic performance of a VAV box. In reality, temperature setpoint control can be more complex. For example, under ASHRAE 36~\cite{ASHRAE36}, control sequences at the air handling unit combine with control sequences at the VAV systems to determine temperature in the conditioned space. We note that this guideline was introduced in 2018 but has not been widely adopted yet. For example, the ASHRAE guidelines recommend using pressure reset, one type of energy-efficient control strategy; however, in our previous experimental work we did not observe pressure reset changes during load shifting~\cite{SHIFDR, AustinACEEE}. Future research should take into account ASHRAE 36 guidelines to study how more energy-efficient controls affect load shifting performance. Doing so would require adapting our building model (red box in Fig.~\ref{fig:controldiagram}) to account for energy-efficient control guidelines but does not require changes to our proposed load shifting control strategies. 

\subsubsection{Power control}
\label{sec:powercontrol}
The power controller wraps around the temperature controller and adjusts the temperature setpoint to match the fan power to a desired reference. These additional control elements are shown in blue in Fig.~\ref{fig:controldiagram}. Unlike in~\cite{Lin_2R2C_model_paper}, we are not overriding the temperature setpoint controller to achieve power control. This method views the building and its existing controllers (shown within the red box in Fig.~\ref{fig:controldiagram}) as a black-box system with an input (temperature setpoint) and output (fan power). 

We run a simulation with only the temperature setpoint controller to determine the baseline fan power, which is shown as ``Baseline Power" in Fig.~\ref{fig:controldiagram}. We subtract the baseline fan power from the measured fan power $P_\mathrm{f}$ to give $P_\mathrm{f,diff}$, which is what we are trying to control. We use another PI controller, labeled ``Power PI," to track the desired ``Event Power" via temperature setpoint adjustments. When the power controller is engaged, adjustments are added to the open-loop temperature setpoint. The value for the proportional gain is $ k_\mathrm{p,pow} = 3.33 \times 10^{-3}$ and the integral gain is $ k_\mathrm{i,pow} = 2.083\times 10^{-5}$; these were selected to satisfy an energy-neutrality criterion, as described in the following subsection. 

\subsection{Analyzing load shifting events}
\label{sec:analysismethods}

As shown in Fig.~\ref{fig:ExamplePower}, we define $t_\mathrm{start}$ as the start time of the event, i.e., when the power is first requested to change, $t_\mathrm{end}$ as the end time of the event, i.e., when power is no longer requested to change, and $t_\mathrm{settle}$ as the time when the building is assumed to have returned to normal operation. In previous experimental studies, $t_\mathrm{settle}$ was assumed to be 1 hour after $t_\mathrm{end}$~\cite{Adi2019, AustinPESGM}. For this paper, we run the simulation until $t_\mathrm{settle} = 9.72 \text{ hr (35,000~s)}+ t_\mathrm{start}$ to ensure the building has completely settled.

To quantify the change in fan power and fan energy caused by the event, we calculate the round trip efficiency (RTE) of the virtual battery, first defined in \cite{BEIL_EFF_FAST_DR}. Specifically, the charging energy $E_\mathrm{in}$ and discharging energy $E_\mathrm{out}$ are
\begin{eqnarray}
    E_\mathrm{in} &=& \int_{t_\mathrm{start}}^{t_\mathrm{settle}} \max \left[ P_\mathrm{f,diff}(t), 0\right]dt,
    \label{eqn:Ein}\\
    E_\mathrm{out} &=& -\int_{t_\mathrm{start}}^{t_\mathrm{settle}} \min \left[P_\mathrm{f,diff}(t), 0\right]dt.
    \label{eqn:Eout}
\end{eqnarray}
The RTE is calculated as the ratio of these values:
\begin{equation}
    \text{RTE} = \frac{E_\mathrm{out}}{E_\mathrm{in}}.
    \label{eqn:RTE}
\end{equation}
For a conventional battery $E_\mathrm{out}$ will always be less than $E_\mathrm{in}$. For virtual batteries, there are sometimes cases in which $E_\mathrm{out} > E_\mathrm{in}$, resulting in an RTE larger than unity. Physically, this means the fans consumed less energy than they otherwise would have consumed, meaning the building became more efficient than under normal operation and/or their were impacts on building services such as temperature and ventilation.

Ideally, load shifting does not change the total energy consumption of the fans across the event window $[t_\mathrm{start}, t_\mathrm{end}]$; otherwise it would not be load shifting but instead load shedding (or increasing). Therefore, we study only events that are ``energy neutral," i.e., events for which energy consumption in $[t_\mathrm{start}, t_\mathrm{end}]$ is approximately the same as the energy consumption of the baseline in the same time window. We adopt the criterion for energy-neutral events from~\cite{AustinPESGM}:
\begin{equation}
    \left |\int_{t_\mathrm{start}}^{t_\mathrm{end}} [P_\mathrm{f,diff}(t) ]dt \right | < \alpha,
    \label{eqn:alpha_level}
\end{equation}
where $\alpha$ is a tolerance value. In this work, we use $\alpha=0.05(E_\mathrm{in}+E_\mathrm{out})$. This ensures that (nearly) all of the inefficiency observed is attributed to building settling after the event, i.e., in $[t_\mathrm{end}, t_\mathrm{settle}]$. Open-loop load shifting events are tuned to be energy neutral by manually adjusting the temperature setpoint changes. For closed-loop load shifting events, the control gains $ k_\mathrm{p,pow}$ and $ k_\mathrm{i,pow}$ were tuned such that, for all testing conditions, the power controller would track the power reference signal closely enough to satisfy~\eqref{eqn:alpha_level}.

To quantify the effect on the building and occupants we use the room temperature root mean squared error (RMSE),
\begin{equation}
    \text{RMSE}_\mathrm{temp} = \sqrt{\frac{\int^{t_\mathrm{settle}}_{t_\mathrm{start}}\left(T_\mathrm{r}(t) - T_\mathrm{r,base}(t) \right)^2dt}{t_\mathrm{settle} - t_\mathrm{start}}},
    \label{eqn:RMSE}
\end{equation}
where $T_\mathrm{r}(t)$ is the measured room temperature during an event and $T_\mathrm{r,base}(t)$ is the baseline room temperature if no event had occurred. 

\section{Simulation results}
\label{sec:results}

In this section, we discuss our findings from simulating the mixing air model. First, we show that results using the mixing air model match experimental results more closely than the original RC model from~\cite{Lin_2R2C_model_paper}. Next, we show a link between changes in mixing and virtual battery efficiencies. Finally, we demonstrate how adding closed-loop control to force settling of the fan power may improve virtual battery performance.

\subsection{Comparison to experimental results}

We use the 2021 data from the SHIFDR dataset~\cite{SHIFDR} to qualitatively compare results from the mixing air model to experimental results. This dataset contains five years of data from 14 buildings in southeast Michigan where over 900 open-loop GTA load shifting events were performed in the summer months. 

\begin{figure}
    \centering
    \includegraphics[width=\linewidth]{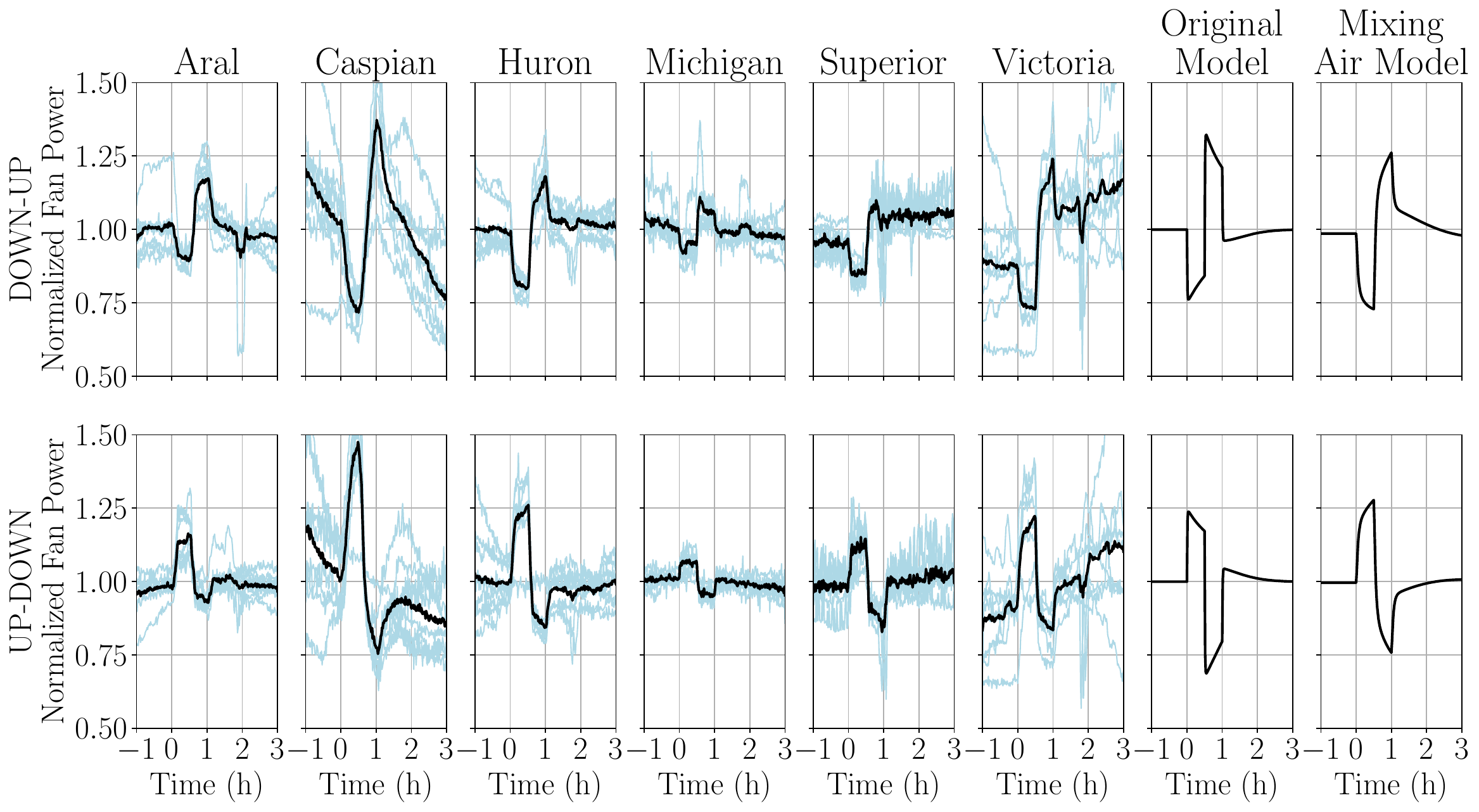}
    \caption{Comparison of 2021 experimental data from SHIFDR~\cite{SHIFDR}, the original model developed in~\cite{Lin_2R2C_model_paper}, and the new mixing air model developed in this paper. For the experimental data (first six sets of plots), we plot the individual events in blue and their time-series average in black. Qualitatively, the mixing air model better matches the experimental data than the original model.}
    \label{fig:comparePower}
\end{figure}

As seen in Fig.~\ref{fig:comparePower}, the mixing air model matches experimental results better than the original model presented in~\cite{Lin_2R2C_model_paper}. The figure compares the fan power obtained from simulations of the original and new models to the fan power measured in six real buildings (anonymized using the names of large lakes) during both DOWN-UP and UP-DOWN load shifting events. All experimental and simulated events employ open-loop GTA with temperature setpoint changes of $\pm1$°F (0.6°C). For the mixing air model, we use $r = 0.3$ and $c = 0.1$. Fan power for each event and each building is normalized such that the average power across the plotted window is unity. The events shown in Fig.~\ref{fig:comparePower} are not necessarily energy neutral.  

A key feature of the experimental data is a two-part response to changes in the temperature setpoint. A clear example of this behavior is observed in the building ``Huron" where, after each temperature setpoint change, the fan power makes a large initial change followed by a slower drift in the same direction as the initial change. In the original model, the slower drift in fan power is in the opposite direction as the initial change, failing to match the experimental results. The poor virtual battery efficiency measured from the experimental data is, in part, a byproduct of most buildings' slow return to normal operation, which is shown by this two-part response. The original model does not capture this behavior and overestimates the buildings' efficiency. In contrast, the mixing air model more-accurately captures the two-part settling response and produces results that appear qualitatively more similar to the experimental data.

\subsection{Link between mixing and efficiency}

\begin{figure}
    \centering
    \includegraphics[width=\linewidth]{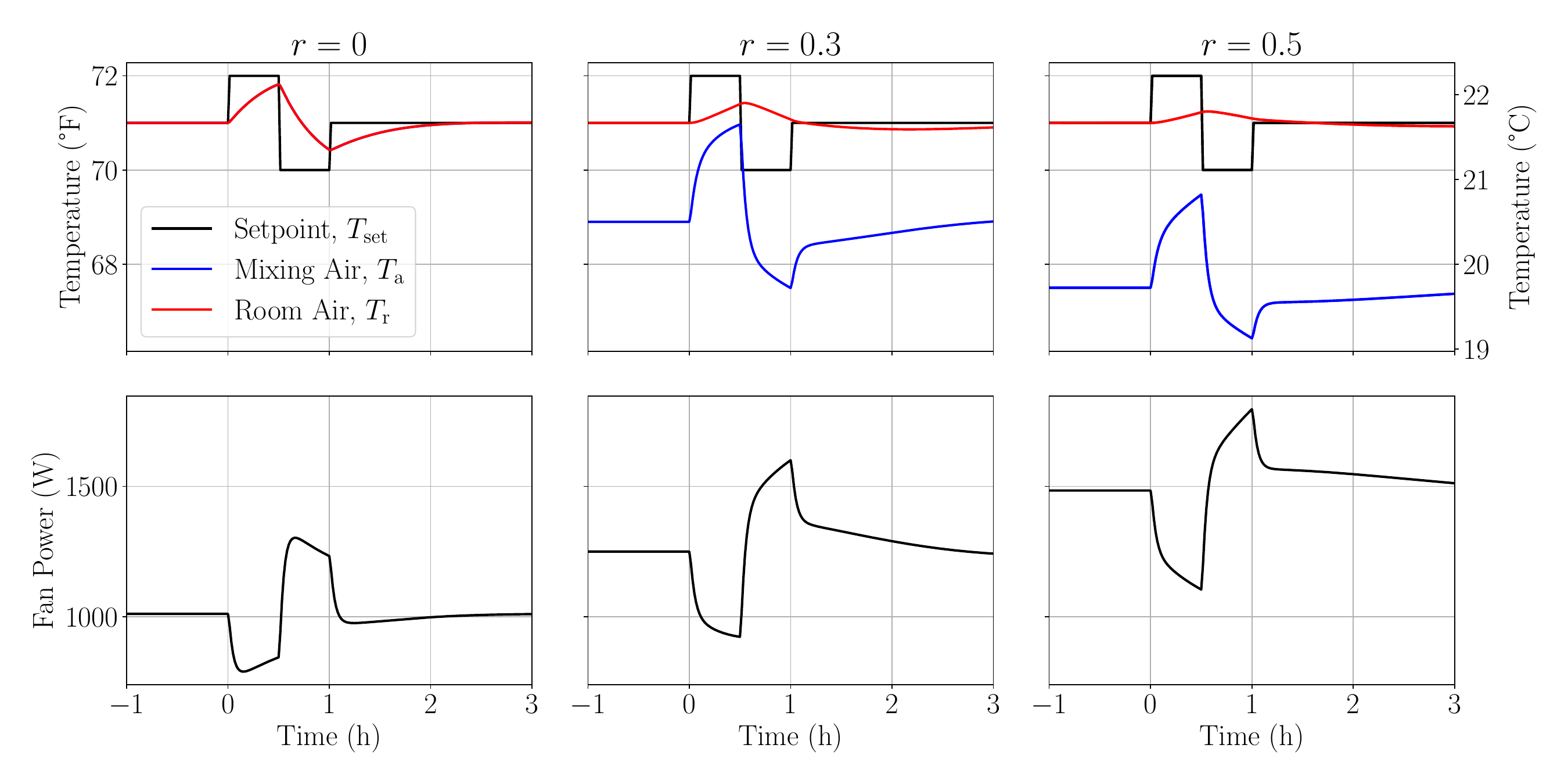}
    \caption{Temperature and fan power trajectories for different relative thermal resistance $r$ values for the mixing zone during an open-loop DOWN-UP event.}
    \label{fig:ExampleThermalResistance}
\end{figure}

By adjusting the parameters $r$ and $c$, we can investigate the extent to which incomplete mixing affects virtual battery inefficiency. Example simulation results for varying values of $r$ are shown in Fig.~\ref{fig:ExampleThermalResistance}. We show temperature and fan power trajectories for an open-loop UP-DOWN event with $c=0.1$. When $r=0$ the mixing zone temperature is the same as the room temperature, indicating that the supply air is well mixed into the room. For $r>0$, the mixing zone temperature is below the room temperature, representing a temperature gradient in the room. The fan power also increases, as the supply air becomes less effective at exchanging heat with the room. The two-part response of the fan power becomes more exaggerated with larger $r$, as it takes longer for the supply air to change the room temperature. The initial large step is caused by the proportional controller reacting to the new setpoint, and the slower drift is caused by the integral controller slowly accumulating error.

\begin{figure}
    \centering
    \includegraphics[width=\linewidth]{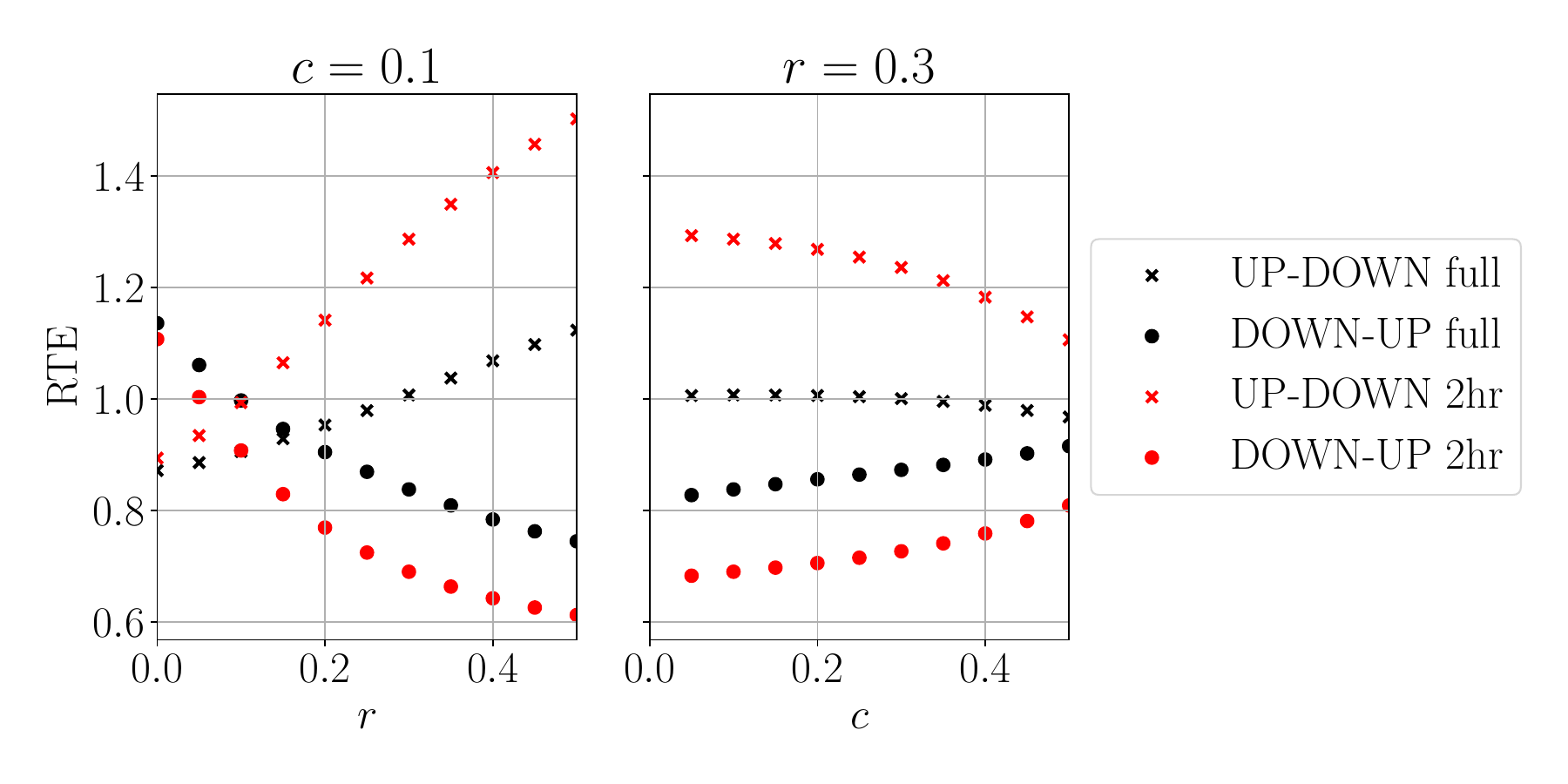}
    \caption{RTE versus $r$ (left) and RTE versus $c$ (right) computed with the full settling window, $t_\mathrm{settle} = 9.72$~hr (black), and computed assuming  $t_\mathrm{settle} = 2$~hr (red).}
    \label{fig:RTEvsMixing}
\end{figure}

Next, we simulated closed-loop energy-neutral events for varying values of $r$ an $c$. Events were designed to increase and decrease the nominal fan power by 10\%. The building was returned to temperature setpoint control at the event conclusion ($t_\mathrm{end}$). The RTEs associated with each event are shown in Fig.~\ref{fig:RTEvsMixing}. We calculate the RTE using the full settling window ($t_\mathrm{settle} = 9.72$~hr, as explained in Section~\ref{sec:analysismethods}) and for $t_\mathrm{settle} = 2$~hr, which is the value commonly assumed in past research. As seen in the left plot in Fig.~\ref{fig:RTEvsMixing}, increasing the value of $r$ initially makes the event efficiency better and then makes the event efficiency worse, indicated by the RTE being further from unity. We see a similar relationship between RTE and $r$ as we have seen between the RTE and building response time (which is affected by mixing and other building physics/limits) in previous work~\cite{AustinPESGM}. As seen in the right plot in Fig.~\ref{fig:RTEvsMixing}, there is less change in RTE as $c$ increases; however, the RTE might be slightly improving. In both cases, by not allowing the building to fully settle after the event (i.e., using a 2~hr settling window), the perceived efficiency is worse, which may have biased past results. These results also explain why, for some buildings, UP-DOWN results are more efficient and, for others, DOWN-UP events are more efficient; this may be a function of their level of mixing.

\subsection{Forced settling via closed-loop control}

\begin{table*}[t]
    \centering
    \caption{RTE and RMSE for cases considering varying baseline error ($r = 0.5$ and $c = 0.3$)}
    \begin{tabular}{c|cc|cc|cc|cc|cc}
    \hline
         & \multicolumn{2}{c|}{Unforced} & \multicolumn{2}{c|}{Forced}& \multicolumn{2}{c|}{Case 1}& \multicolumn{2}{c|}{Case 2}& \multicolumn{2}{c}{Case 3}\\
      Event   & RTE    & RMSE (°F) & RTE    & RMSE (°F)& RTE    & RMSE (°F) & RTE    & RMSE (°F) & RTE    & RMSE (°F) \\

    \hline
   UP-DOWN & 1.1444  &  0.0358  &  0.9072  &  0.0227 &0.9033   & 0.0226  &  0.9595   &  0.0243  &  0.8567  &  0.0215\\
   DOWN-UP &  0.7526  &  0.0362  &  0.9884   & 0.0228 & 0.9897  &  0.0228  &  1.0515   & 0.0223  &  0.9315 &   0.0240\\
   \hline
    \end{tabular}
    \label{tab:ForcedSettling}
\end{table*}

To improve settling, and thus RTE, we can guide the fan power back to the baseline fan power before returning control back to the temperature setpoint controller. We add a 1-hour period after the event ends where the fan power controller is still active, but the event power is 0~W. We refer to this control as ``forced settling," as we use the power controller to force the building back to baseline operation. This type of control is dependent on the accuracy of the baseline model, as an incorrect baseline would cause the controller to push the building temperature and fan power away from the states that would eventually be achieved through normal operation (i.e., temperature setpoint control). In this case, after one hour of forced settling, the building would not be settled, but would need to settle these states.

\begin{figure}
    \centering
    \includegraphics[width=\linewidth]{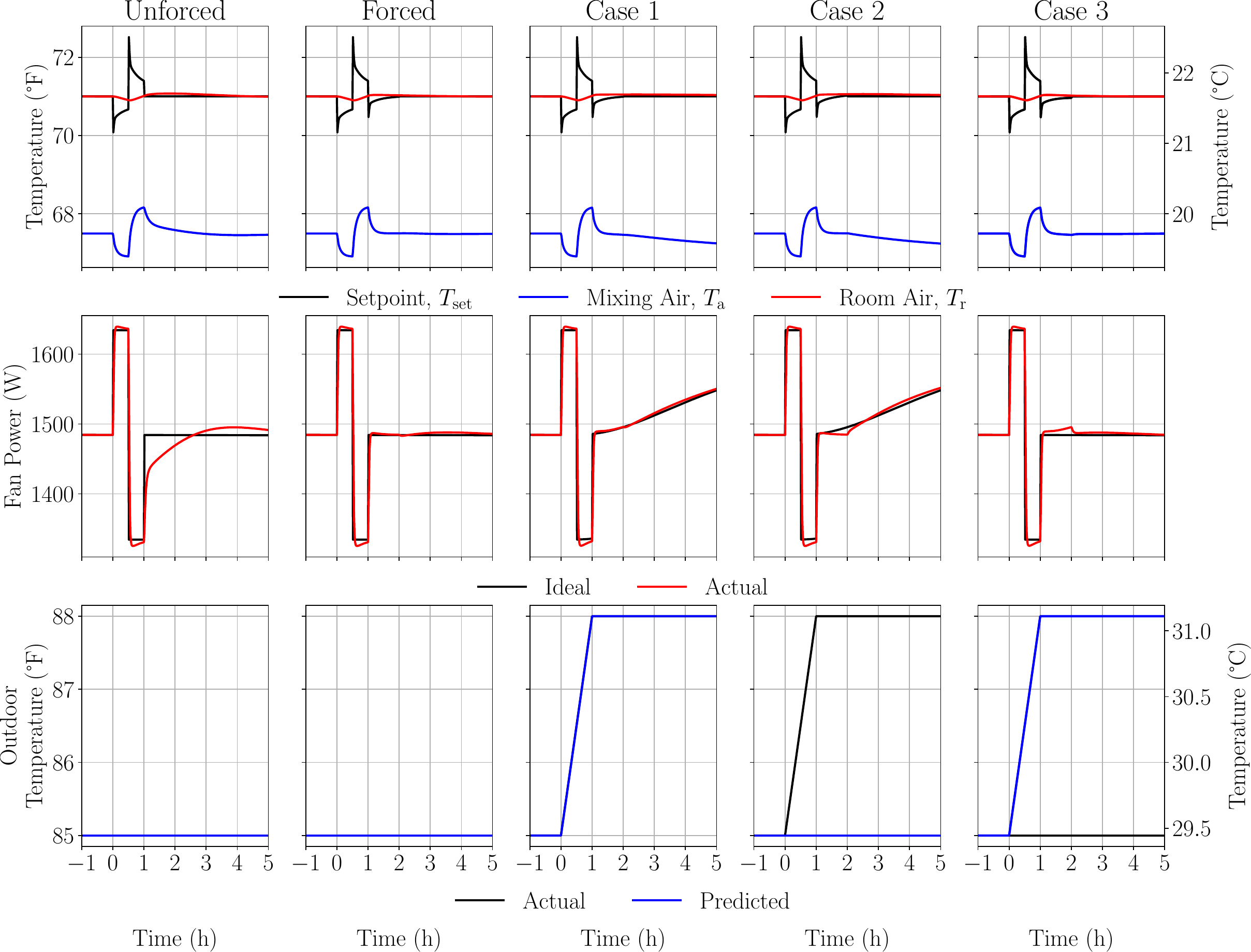}
    \caption{Temperatures (top), fan powers (middle), and outdoor temperatures (bottom) for forced and unforced settling, when $r=0.5$ and $c=0.3$. Cases 1-3 show forced settling where 1) outdoor conditions change and the baseline model predicts the change, 2) outdoor conditions change but the baseline model does not predict the change, and 3) outdoor conditions do not change but the baseline model predicts a change.}
    \label{fig:ForcedExample}
\end{figure}

Fig.~\ref{fig:ForcedExample} compares unforced and forced settling. The top plots show the temperatures and temperature setpoints. The middle plots show the actual fan power and ideal fan power for a perfect load shift. The bottom plots show the actual and predicted outdoor temperature, which is used to calculate the baseline. The ``unforced case" returns to temperature setpoint control immediately after the event concludes while the ``forced case" includes the 1-hour forced-settling period. Cases 1, 2, and 3 represent situations in which the actual and/or predicted outdoor conditions change; these cases will be discussed later in this section. The RTE and room temperature RMSE for all scenarios in Fig.~\ref{fig:ForcedExample} are included in Table~\ref{tab:ForcedSettling}. 

As shown in the first two plots in the middle row of Fig.~\ref{fig:ForcedExample}, the fan power in the unforced case takes longer to settle than in the forced case. In the forced case, we see a dynamic temperature setpoint after the event ends to guide the fan power back to the baseline. There is a small deviation in fan power when the temperature controller resumes at the 2-hour mark; however, this is minor. As seen in Table~\ref{tab:ForcedSettling},
forced settling also improves the efficiency of the event, i.e., the RTEs are closer to unity (i.e., 0.9072 vs.\ 1.1444\footnote{Note that in Fig.~\ref{fig:ForcedExample}, in the unforced case, the building is warmer than desired after the event, i.e., the building uses less energy (RTE $>1$) but delivers less service. In the forced case, this is corrected (the building delivers approximately the requested service) at the expense of more energy use, leading to an efficiency less than 1 but still closer to 1 than in the unforced case. This could be perceived as a reduction in efficiency but instead should be seen as correcting the control to deliver the requested service.} for UP-DOWN and 0.9884 vs.\ 0.7526 for DOWN-UP). Forced settling also provides less disruption to the occupants as indicated by a smaller room temperature RMSE in the forced case than in the unforced case. Therefore, through closed-loop control and an accurate estimate of the baseline fan power, the load shifting inefficiency and effect on the building can be reduced. We note that, with this type of control, the RTE will never be exactly unity, as there are physical limits on how closely the building can follow a power signal.

We also explore how actual and/or predicted changes in outdoor conditions affect forced settling, by varying the outdoor air temperature during the event. As discussed in Section~\ref{sec:powercontrol}, the baseline fan power (used by the fan power controller) is computed by simulating the building with no event. This requires a prediction of the outdoor air temperature. Previously, we assumed that the outdoor air temperature during simulated events was constant and exactly matched the predicted temperature. We now consider three new cases in which one or both of these assumptions is no longer true. In Case 1, the actual outdoor temperature changes and the predicted outdoor temperature accurately matches the change. In Case 2, the actual outdoor temperature changes, but the predicted outdoor temperature remains constant. In Case 3, the predicted outdoor temperature changes, but the actual outdoor temperature remains constant. For all 3 cases we consider a 3°F (1.7°C) increase in outdoor temperature. Power and temperature trajectories for these cases are shown in Fig.~\ref{fig:ForcedExample}, while RTE and room temperature RMSE are included in Table~\ref{tab:ForcedSettling}.

Case 1 has similar RTE and room temperature RMSE as in the forced settling case. This indicates that, with a good prediction of the baseline fan power, closed-loop control can perform well under changing outdoor conditions. Cases 2 and 3 also have similar room temperature RMSE as in the forced settling case. This indicates that the occupant comfort will be similar, despite inaccurate baseline fan power prediction. However, the RTE is worse in Case 3 than in Cases 1 or 2, or the forced settling case, indicating that incorrectly predicting changes in outdoor conditions may be worse than predicting no change at all. We postulate that the large thermal mass of the building reduces the effects of changing outdoor conditions and provides a buffer for the power controller during forced settling. We also note that an accurate prediction of outdoor conditions is only one aspect of accurate baseline prediction. In practice, occupancy and other time-varying human factors affect building energy usage, and these complex impacts are usually captured in baseline models in overly-simple ways. This would present more challenges for our closed-loop control approach. More research is needed to explore the impacts of these challenges.

\section{Conclusion}
\label{sec:conclusion}

In this paper, we developed and analyzed a new analytical building model to demonstrate how variables not typically considered by building automation and control systems affect the efficiency of a virtual battery capturing the dynamics of HVAC fans. Specifically, we consider the effects of incomplete mixing of supply air into the conditioned space during sub-hourly load shifting of HVAC fans. We found this model produces simulation results that qualitatively match experimental results better than previous models. Our results show that as the mixing of supply air becomes worse, so does the efficiency of the virtual battery. This presents challenges to grid operators who wish to model buildings as virtual batteries, as air mixing is difficult or impossible to measure. We note that incomplete mixing is only one example of building phenomena that are unmeasured during building operation. Other phenomena, e.g., related to air pressure or reheat, may also impact the virtual battery efficiency and go unmeasured, leaving the grid operator unsure of the virtual battery efficiency.

However, we showed that closed-loop control of the fan power could mitigate this issue without the need for accurate measurements of air mixing. Specifically, a feedback loop from the fan power to the temperature setpoint can reduce inefficiency from load shifting. This method requires accurate prediction of the baseline fan power and direct measurement of the fan power, a quantity not usually measured. However, this could be achieved through fan power submetering. We found that imposing a forced settling period after the event significantly improves the virtual battery efficiency. Furthermore, inaccurate baseline predictions may not have a large impact on the performance of this method, and forced settling performed better than unforced settling, in all cases in which there was baseline error. However, it is still unclear if larger errors in the fan power baseline would provide similar results. Future research should explore the performance of closed-loop control with poor baseline predictions to determine the sensitivity of control performance to prediction error.

\bibliographystyle{ieeetr}
\bibliography{ref.bib}

\end{document}